\input{epsf}
\documentstyle[prb,aps,multicol,amssymb]{revtex}
\renewcommand{\narrowtext}{\begin{multicols}{2} 
\global\columnwidth20.5pc} \renewcommand{\widetext}{\end{multicols} 
\global\columnwidth42.5pc} \multicolsep = 8pt plus 4pt minus 3pt

\begin{document}
	\draft

\title{Anomalous dielectric after-effect in ferroelectric 
KH$_2$PO$_4$}
\author{J. Gilchrist}
\address{Centre 
de Recherches sur les Tr\`es Basses Temp\'eratures, laboratoire 
associ\'e \`a l'Universit\'e Joseph Fourier, CNRS, BP 166, 38042 
Grenoble Cedex 9, France}
\maketitle

\begin{abstract}
Dielectric permittivity, 
$\epsilon=\epsilon^\prime-i\epsilon^{\prime\prime}$ of KH$_2$PO$_4$ 
pressed powders was measured between $T$~= 1.4 and 25~K in the 
presence of $dc$ electric bias fields. Usually if the bias was switched at time 
$t$~= 0, $\epsilon^\prime$ and $\epsilon^{\prime\prime}$ jumped to 
new values then decreased approximately as $\log t$ (6~s~$<t<2000$~s). 
This well-known effect, that is also found with single 
crystals, is attributed to switchable microdomains that are present 
accidentally in crystals but are systematically more numerous in 
powders. A very different after-effect was observed in a narrow $T$ 
range around 7--8~K. $\epsilon^\prime$ jumped to a lower value then 
increased with $t$ according to a stretched exponential with a 
$T$-dependent time 
constant. This lay near the extrapolation of the Arrhenius law of a known, 
but unassigned weak-field dispersion that is a property of polydomain 
single crystals as well as powders. The weak-field dispersion is 
attributed to the elementary movement of a jog on a 
lateral step displacement of a domain wall, consisting of a single 
H-bond reversal. The anomalous after-effect results from the 
interaction between these point defects and the microdomain system.
\end{abstract}


\narrowtext

\section{Introduction}\label{intro}

\begin{figure}
\centerline{\epsfxsize=7.5 cm \epsfbox{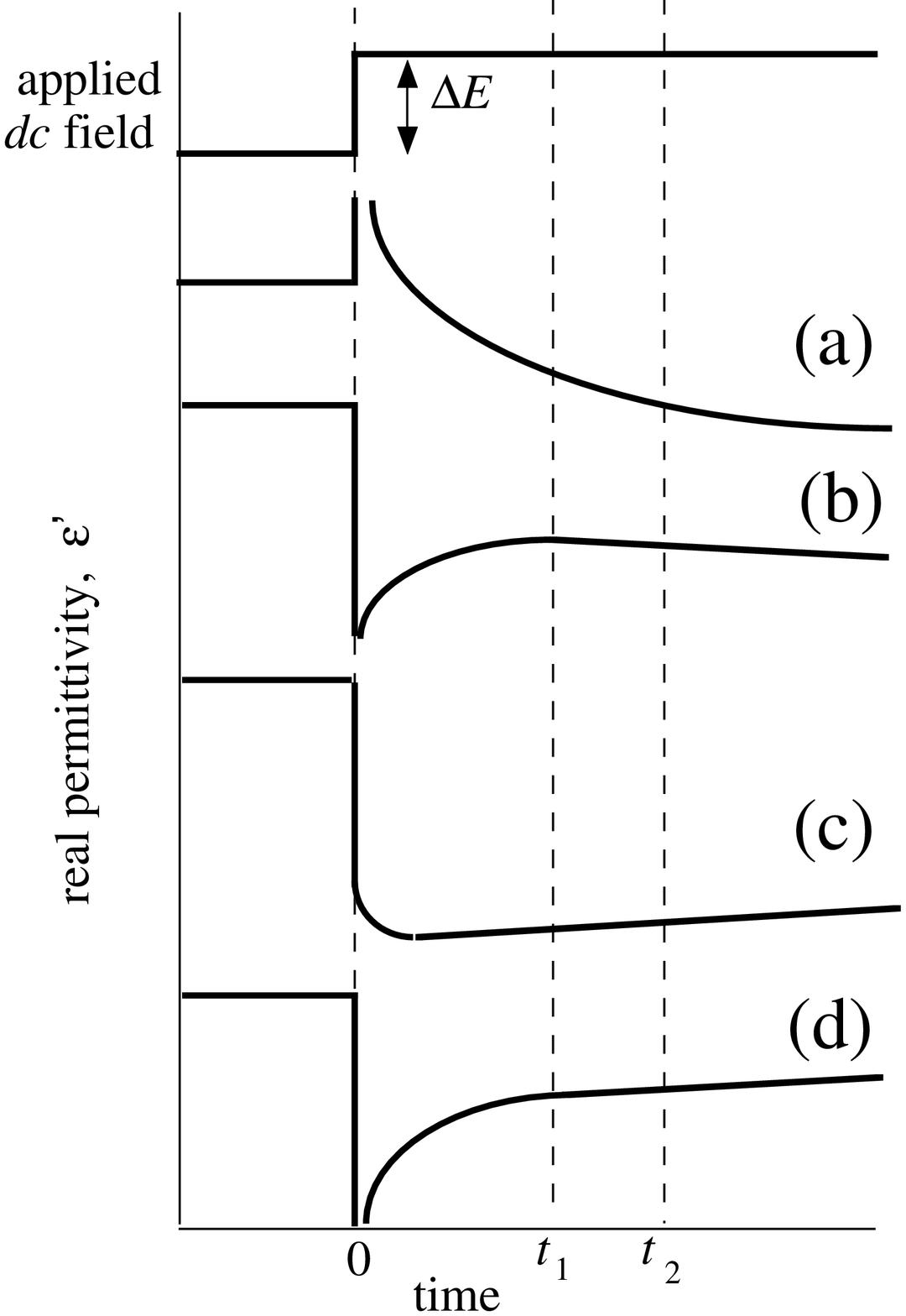}}
\caption{Schematic illustration of dielectric after-effects. The field 
applied to the sample consists of an $ac$ measuring field of fixed low 
amplitude plus a $dc$ bias field. After the sample has been stabilised 
at a given temperature for some time, the bias field is switched. The 
real (in-phase) part of the permittivity, $\epsilon^\prime$, changes 
abruptly then usually relaxes downwards (a). It may also relax 
upwards at short times (b), at long times (c), or at all times 
accessible to experiment (d). A slope $s^\prime=d\epsilon^\prime/d\ln t$ 
is defined as the best fit to the data within a specified range 
$t_1<t<t_2$. This is particularly useful for cases like (a) where 
$s^\prime$ takes a negative value that only depends weakly on $t_1$ 
and $t_2$. The 
imaginary (quadrature) permittivity behaves similarly but on a reduced scale, 
and $s^{\prime\prime}=d\epsilon^{\prime\prime}/d\ln t$.}
\label{fig1}
\end{figure}

It has been known\cite{BarklaPM53} since the early days of 
ferroelectric KH$_2$PO$_4$ (KDP) that the material exhibits a retarded 
response to a disturbance as well as a prompt one. In particular, if 
its dielectric permittivity, 
$\epsilon=\epsilon^\prime-i\epsilon^{\prime\prime}$, is recorded using 
a low amplitude $ac$ field in the additional presence of a $dc$ bias field 
that is changed abruptly at time $t=0$, then $\epsilon^\prime$ 
responds by promptly taking a new value, then by decreasing with $\log 
t$ as shown in Fig.~\ref{fig1}, curve (a). 
Zimmer, Engert, and Hegenbarth\cite{ZimmerFL87} reported 
this for a single crystal with the fields parallel to the 
ferroelectric $c$-axis, at temperatures $T$~= 4.2, 20.4, 77.8, and 
300~K. The property is shared by many 
disordered ferroelectrics and will be referred to as ``the 
normal after-effect''.

Part of the interest of such studies lies in comparison with 
dielectric glasses which also exhibit a ``normal after-effect'' 
at $T<1$~K.\cite{SalvinoPRL94} It is also well 
known that ferroelectrics with diffuse transitions and relaxor 
ferroelectrics display low 
temperature thermal properties somewhat similar to structural 
glasses.\cite{DeYoreoPRB85,HegenbarthF95} A $T^{3/2}$ specific heat 
term at $T<5$~K 
was also at one time reported for KDP\cite{LawlessPRL76,LawlessPRB76} 
though KDP has a normal, sharp ferroelectric transition. It was later 
clarified that large, pure KDP crystals can have almost Debye-like 
specific heat,\cite{LawlessF87} and lack a ``glassy'' thermal 
conductivity anomaly.\cite{DeYoreoPRB85}

The present article reports a study of the dielectric after-effect in 
KDP, mostly as pressed powders. These would be certain to have a 
strong $T^{3/2}$ 
specific heat term, and the after-effects were found to be stronger, 
and less variable from one sample to another than with single 
crystals. In the course of this study, a curious anomaly was 
found.\cite{GilchristSSC98} In a narrow temperature range around 
7.5~K the effect of a bias switch was a prompt decrease of 
$\epsilon^\prime$ followed by an upward relaxation 
according to a stretched exponential law. Curves (b), (c) and (d) in 
Fig.~\ref{fig1} show schematically what was observed at 9.0, 6.0 and 
7.5~K respectively. The retarded response appeared to be the sum of 
two terms, the usual downward relaxation as $\log t$ (the normal 
after-effect) and an upward relaxation with a characteristic time 
constant that depended on $T$. Near 7.5~K, the upward relaxation 
dominated throughout the range 6~s~$<t<2000$~s, at 9.0~K only at the 
shorter times, at 6.0~K the longer ones. This term will 
be referred to as ``the anomalous after-effect''. Two electroded single 
crystal samples exhibited the anomalous after effect, as well as the 
normal one,\cite{GilchristSSC98} but both effects were more 
precisely measurable using pressed powders.

Among many previous dielectric studies of KDP, ones by Holste, 
Lawless and Samara\cite{HolsteF76} and by Motegi, Kuramoto, and 
Nakamura\cite{MotegiJPSJ83,KuramotoJPSJ84} paid particular attention 
to temperatures below 25~K. Over a continuous background of absorption 
and dispersion rising regularly with $T$, Motegi {\it et al.} observed 
two specific dispersions, one sensitive to fields parallel to the 
ferroelectric axis ($E\parallel c$), that required the presence of 
domain walls, the other sensitive to $E\parallel a$ and equally 
present in poly- or mono-domain crystals. The $E\parallel c$ 
dispersion (hereafter ``KMN-C'') obeyed an Arrhenius law with 
activation energy $A$~= 19~meV and pre-exponential frequency, $f_0$~= 
45~GHz,\cite{KuramotoJPSJ84} i.e., it was centered at frequency 
$f=f_0\exp(-A/kT)$. It looks like the typical effect of a reorientable 
point defect species in a crystalline environment. A link between 
the anomalous after effect 
and KMN-C was suggested\cite{GilchristSSC98} by the 
finding that the time constant of the stretched exponential obeyed the 
same Arrhenius law as KMN-C. As it will be shown in 
Sec.~\ref{singlecrystals} below, the agreement is not as exact as 
initially supposed, but it will also be reported (Sec.~\ref{particlesize}) 
that the strengths of the two effects were closely correlated, when 
samples of different qualities were compared.

The main aim of the present article is to report the anomalous after 
effect and to suggest an explanation for it. Since it appears to be 
related both to the normal after effect and to KMN-C, a report and 
discussion in some detail of each of these effects is also necessary.

It is notorious that a dielectric study of any solid sample that is 
not a properly electroded single crystal can lead to spurious results 
that bear little relation to bulk material properties. It is argued 
below in Sec.~\ref{results} and \ref{particlesize} that serious errors 
are avoided by restricting attention to $T<25$~K, and interpreting 
the measurements cautiously. The general features of the data are 
also described in Sec.~\ref{results}, by reference to one 
representative KDP sample. In Sec.~\ref{behavior} the normal 
after-effect is reported. This allows scaling parameters to be 
defined, that are useful in the subsequent (Sec.~\ref{T25K} and 
\ref{particlesize}) description of the anomalous after-effect. Results 
obtained with four single crystal samples are reported in 
Sec.~\ref{singlecrystals}, and the three effects, their interrelations 
and possible origins are discussed in Sec.~\ref{discussion}.

\section{Samples and measurements}\label{measurements}

The pressed powder samples were derived from Aldrich 99+\%~ACS 
reagent. In most cases the product was ground 
manually to $\sim~3~\mu$m particle size (as examined by swept-beam 
electron microscopy at 20~kV) then pressed ($\sim~50$~MPa) into 
pills of diameter 10~mm and thickness normally 250--400~$\mu$m, 
density~$>$~80\% of crystal density. The pills were pressed 
($\sim~100$~kPa) between lead or indium electrodes. Four single 
crystal samples were also 
studied, and these are described in Sec.~\ref{singlecrystals}, where 
their results are also reported.

Capacitance and 
conductance were measured using a General Radio 1621 transformer 
bridge system, and expressed as $\epsilon^\prime$ and 
$\epsilon^{\prime\prime}$ based on the external dimensions of the 
pills and electrodes. The variations of $\epsilon^\prime$ and 
$\epsilon^{\prime\prime}$ during a 360~s interval were obtained from 
a chart recording of the bridge off-balance signals. At longer times 
the bridge was rebalanced for each point. The temperature was controlled 
using a carbon resistor sensor (Allen Bradley 390~$\Omega$), that was 
calibrated periodically by substituting 
a Pt and a Ge resistor for the sample capacitance. Thermometry errors 
were of two types, a short and a long term error. The short term error 
was caused by the carbon resistor varying with time following a 
temperature change. This was noticeable below 10~K and 
non-negligible below 6~K. It was allowed for by assuming that
\begin{eqnarray*}
\frac{d\epsilon^\prime}{dt}=\left(\frac{\partial\epsilon^\prime}{\partial 
t}\right)_T+\left(\frac{\partial\epsilon^\prime}{\partial 
T}\right)_t\frac{dT}{dt},
\end{eqnarray*}
where $d\epsilon^\prime/dt$ was measured and 
$(\partial\epsilon^\prime/\partial t)_T$ was required. By waiting long 
enough before switching the bias field, the last term became a slow 
drift that could be subtracted confidently. The ``long term'' error 
had various contributing causes. In routine work with different 
samples, the absolute $T$ was known to $\pm 0.5$~K, but for the 
extensive work on Sample~2, to $\pm 0.2$~K. For the detailed study around 
7.5~K (Fig.~\ref{fig9}) the relative error was $\pm 0.05$~K (rapid 
succession of measurements without heating above 122~K). For 
Fig.~\ref{fig12}, absolute $T$ was known to $\pm 0.01$~K.

\section{Results, Generalities}\label{results}

The results reported here and in Sec.~\ref{behavior} were obtained 
with a typical pressed powder sample (Sample~1) prepared from KDP as 
received (not recrystallized).

\subsection{Weak $ac$ field, no bias}\label{weakfield}

\begin{figure}
\centerline{\epsfxsize=7.5 cm \epsfbox{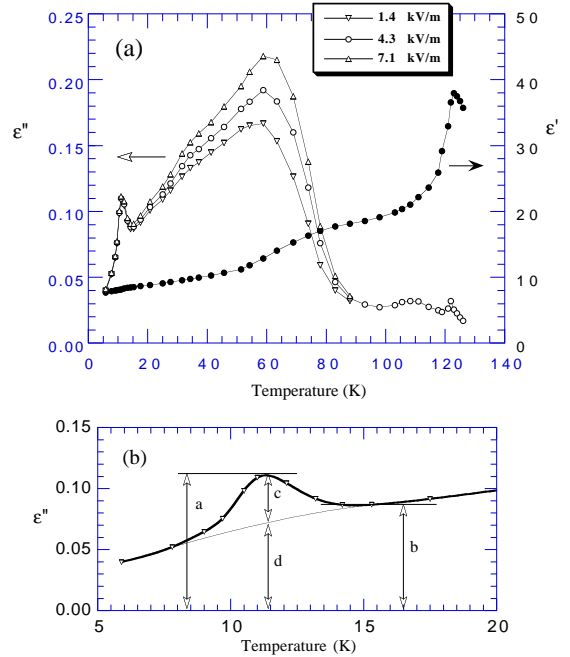}}
\caption{(a) Real and imaginary dielectric permittivity of 
a typical pressed powder sample (Sample~1) at 1.2~kHz and three 
field levels. Here and in Figs.~\ref{fig4}, \ref{fig5}, 
\ref{fig7}, and \ref{fig8}, $\epsilon^\prime$ or related data 
are shown as filled symbols, $\epsilon^{\prime\prime}$ data as 
corresponding outline symbols. The present data were not 
sensitive to temperature history on a scale that would be 
visible on the plot, provided the sample was held at each $T$ value 
for 10 minutes or more. $\epsilon^\prime$ is shown only for 4.3~kV/m 
as the other data would be indistinguishable on the plot. 
(b)~Low $T$ detail ($\epsilon^{\prime\prime}$ at 1.4~kV/m only). 
The strength of the specific absorption, $c$ relative to the background 
might be expressed by the ratio $c/d$, which requires an interpolation 
of the background curve. For comparison of different samples the ratio $a/b$ 
is used.}
\label{fig2}
\end{figure}

Figure~\ref{fig2}
shows the response of Sample~1 to 
alternating fields 
of different strengths and a fixed frequency. At ambient $T$, sample 
impedance was limited by conduction, but this diminished rapidly 
on cooling, and became 
undetectably small already well above the ferroelectric transition 
temperature, $T_c$, and outside the range of Fig.~\ref{fig2}. In 
Fig.~\ref{fig2}, $T_c$ 
is marked by a peak of $\epsilon^\prime$ at the usual value of 122~K. 
Well below $T_c$ there are $\epsilon^{\prime\prime}$ peaks near 12~K 
and near 60~K. 
The 12~K peak is a genuine property of the material but the 60~K 
peak is not. A spurious peak is to 
be expected near 60~K for the following reason. For the sake of 
argument, suppose each grain had an anisotropic permittivity as 
measured with a properly electroded, unclamped single crystal. Both 
$\epsilon^\prime_c$ and $\epsilon^{\prime\prime}_c$ rise rapidly 
with $T$ in this range,\cite{BarklaPM53,MotegiJPSJ83} 
$\epsilon^\prime_c$ changing in order of magnitude from $\approx 10$ 
to $>10^4$. At $\epsilon^\prime_c\approx 10$, there would be 
appreciable, though nonuniform penetration of $E\parallel c$ field 
component into suitably oriented powder grains, but at 
$\epsilon^\prime_c>10^4$ such penetration would be negligible, the 
applied field being confined to the vacuum gaps, and to $E\parallel 
a$ within the grains ($\epsilon^\prime_a$ remains moderate, rising to 
a peak value $\approx 60$ at $T_c$, while $\epsilon^{\prime\prime}_a$ 
remains low). The best compromise between the rising 
$\epsilon^{\prime\prime}_c(T)$ and the falling $E\parallel c$ 
penetration then locates a loss peak as in Fig.~\ref{fig2}(a). It is 
not a Maxwell-Wagner effect and does not involve $dc$ conduction.

On the other hand at $T<25$~K, crystal $\epsilon_a$ and $\epsilon_c$ 
values are both moderate and depend weakly on $T$. At the moderate 
field values used, the dielectric response did not depend markedly on 
field strength (Fig.~\ref{fig2}(a)), while $dc$ conductance was totally 
insignificant. Insofar as the constituents of a composite dielectric 
material can be regarded as continuous media, the $\epsilon$ value of 
the composite is a mean value of the $\epsilon$ of the constituent 
parts. The possible complex mean values lie within rigorously 
determined 
bounds.\cite{BergmanPR78,BergmanPRL80,MiltonAPL80,MiltonJAP81} In 
these conditions, the 12~K loss peak in Fig.~\ref{fig2} can be 
assigned as a peak in $\epsilon_c$ or $\epsilon_a$ or both. In 
absolute value, it is distinctly stronger than either the 
$\epsilon_c$ or $\epsilon_a$ dispersions reported by Motegi {\it et 
al.}\cite{MotegiJPSJ83,KuramotoJPSJ84} in single crystals, but in 
position it is nearer KMN-C (see Sec.~\ref{singlecrystals} below and 
Fig.~\ref{fig12}). The 
background value of $\epsilon^{\prime\prime}$ in Fig.~\ref{fig2} is 
also an order of magnitude higher than the background 
$\epsilon^{\prime\prime}_c$ in Motegi's crystals, so the relative 
strength of the special absorption effect is quite similar. It is to 
be noted that the background $\epsilon^{\prime\prime}_c$ value as well 
as KMN-C depended on the presence of domain 
walls,\cite{KuramotoJPSJ84} so the relative strength could be used for 
comparing the KMN dispersion in samples of different qualities. It 
will be useful to characterize this relative strength for different 
samples by 
expressing the ratio of the peak $\epsilon^{\prime\prime}(T)$ near 
12~K to the minimum near 15~K, the ratio $a/b$ in Fig.~\ref{fig2}(b), 
as a percentage. For Sample~1, $a/b$~= 126\% and in absolute value, 
$c\approx 0.042$.

\subsection{Effects of bias switches}\label{switches}

\begin{figure}
\centerline{\epsfxsize=7.5 cm \epsfbox{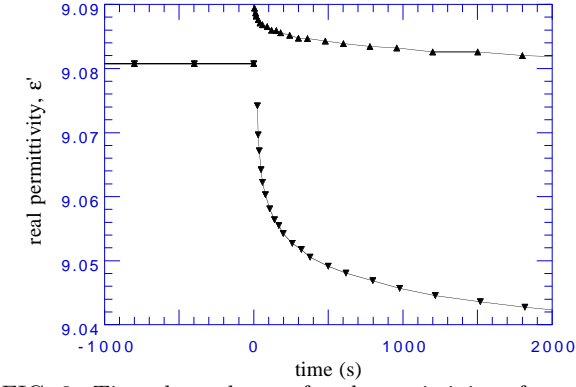}}
\caption{Time dependence of real permittivity of a typical pressed 
powder (Sample~1) at 25~K measured using 4.3~kV/m at 1.2~kHz. Two sets 
of data are shown. 
Points~$\blacktriangle$: the sample was first cooled from above 122~K 
to 25~K in zero bias field and held there for 1~hr, then at $t=0$ a 
bias field of 11.4~kV/m was switched on. Points~$\blacktriangledown$: 
the same, except that the sample was cooled in a bias field of 400~kV/m 
and this field was reversed at $t=0$. $\epsilon^\prime(t)$ depended 
solely on the magnitude (11.4 or 800~kV/m) of the bias step, and not 
on the value (zero or nonzero) of the cooling field.}
\label{fig3}
\end{figure}

Figure~\ref{fig3} shows typically how $\epsilon^\prime$ behaved 
following bias steps of different magnitudes at any $T$ value not near 
7.5~K. ``Step'' here means a 
single abrupt change of applied field. 
Normally after each 
step, and after the subsequent changes of $\epsilon$ had been recorded 
the sample was heated to $T_c$ and recooled. As Fig.~\ref{fig3} 
shows, after a small bias 
step, $\epsilon^\prime$ promptly rose then gradually returned towards 
its original value, but after a larger step it gradually moved to new, 
lower values. A curve like (a) of Fig.~\ref{fig1} would be found after 
a bias step of intermediate magnitude. In each case, taking $t=0$ at 
the bias step, 
$\epsilon^\prime(t>0)\approx\epsilon^\prime(\infty)+Ct^{-p}$, where 
$0<p<0.1$ and $C$ is a constant. To estimate $\epsilon^\prime(\infty)$ 
would require a long, uncertain extrapolation, and over a limited $t$ 
range the power law differs little from a logarithmic variation, so 
it is more useful to define $s^\prime=d\epsilon^\prime/d\ln t$, and 
find $s^\prime$, which was usually a slowly varying function of $t$. 
The variations 
of $\epsilon^{\prime\prime}$ (not shown in Fig.~\ref{fig3}) were 
similar, but scaled down by a factor of $\approx 4$. Defining 
$s^{\prime\prime}=d\epsilon^{\prime\prime}/d\ln t$, $\vert 
s^{\prime\prime}\vert$ was like $\vert s^\prime\vert$, usually a slowly 
diminishing function of $t$. In cases like curves (b) and (c) of 
Fig.~\ref{fig1}, $s^\prime$ depended more strongly on $t$ and even 
changed sign. In every case the final value of $\epsilon^\prime$, for 
$t\rightarrow\infty$, was less than or equal to the initial value 
$\epsilon^\prime(t<0)$. The strength of the $ac$ measuring field was 
unimportant provided it did not exceed $0.4\Delta E$, otherwise $\vert 
s^\prime\vert$ and $\vert s^{\prime\prime}\vert$ were underestimated.

Figure~\ref{fig3} shows two relaxations of $\epsilon^\prime(t)$ that 
start from a same value, $\epsilon^\prime(t<0)$ even though in one 
case the sample had been cooled from $T_c$ in a bias field, the other 
in zero field. At $T\leq 25$~K, $\epsilon^\prime(t<0)$ was generally 
reproduced to within 0.1\% over a series of thermal cycles up to $T_c$ 
and back, even though the bias field was sometimes zero, sometimes 
$\approx 1$~MV/m and the cooling rate also was variable. Similarly the 
value of $\epsilon^{\prime\prime}$ prior to any bias step was 
reproduced to within 1\%. This is quite different from single crystal 
behavior. Three types of bias switches were studied. Either the sample was 
cooled in zero field, and field switched on at $t=0$, or it was field 
cooled and at $t=0$, the field switched off or reversed. Repeated 
measurements at 25~K with bias field switched from 0 to 920~kV/m, 
separated by thermal cycles to $T_c$ yielded standard deviations $\pm$
4\% for $s^\prime$ and $\pm$2\% for $s^{\prime\prime}$, excluding 
data from the first few cycles. If instead the sample was cooled in 
920~kV/m and this bias switched off, $\vert s^\prime\vert$ was 
3$\pm$4\% higher and $\vert s^{\prime\prime}\vert$, 2$\pm$2\% higher. 
These are not significant differences, and the same applies if the 
sample was cooled in 460~kV/m and this switched to -460~kV/m. Only the 
magnitude $\vert\Delta E\vert$ of the bias change was important. In 
this respect also the pressed powders behaved quite differently from 
single crystals. This point established, the three types of bias 
switches were used indifferently. Since also the sign of $\Delta E$ is 
irrelevant, it will be taken to be positive.

A spurious after-effect could have been caused by space charge 
migration. If charge were to gradually migrate and accumulate, the 
effective penetration of the bias field into the grains would 
gradually diminish. This might cause $\epsilon$ to return towards its 
original value, but not to acquire a new, lower value. It will be 
reported (Sec.~\ref{particlesize}) that samples of 
very different qualities always had limiting $s^\prime$ and 
$s^{\prime\prime}$ values of the same order of magnitude, which argues 
against charge migration.

Sequences of bias changes without change of $T$ were not studied 
systematically, but schematically, if the bias field was switched 
periodically between 
values $E_1$ and $E_2$, starting at $t=0$, without changing $T$, each 
switch initiated a new relaxation. In that case 
$\epsilon^\prime(t)\approx\epsilon^\prime(\infty)+C(t-t_i)^{-p}$, 
where $t_i$ is the time of the most recent bias switch. For small 
$\vert E_2-E_1\vert$, 
$\epsilon^\prime(\infty)\approx\epsilon^\prime(t<0)$, while for 
large $\vert E_2-E_1\vert$, after several switches 
$\epsilon^\prime(\infty)$ approached a constant lower value.

\section{Permittivity changes following a bias 
step}\label{permittivity}

A first systematic study of the after-effect at $T\approx 25$~K is 
reported because this is far from any special feature in 
Fig.~\ref{fig2}(a), and the normal after-effect could be observed 
without the anomalous effect.

\subsection{Behavior near 25~K}\label{behavior}

\begin{figure}
\centerline{\epsfxsize=7.5 cm \epsfbox{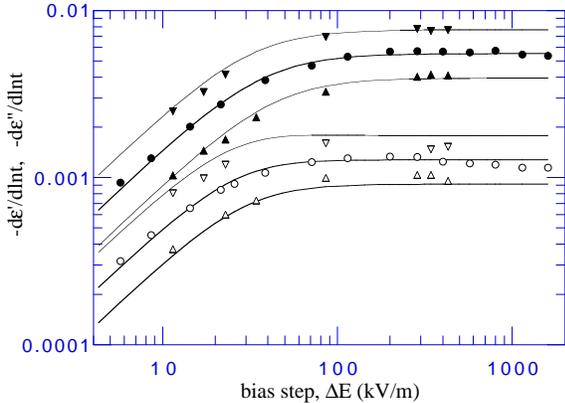}}
\caption{Sample~1 at 25~K in weak $ac$ fields: derivatives of real and 
imaginary permittivity with respect to $\ln t$ where $t$ is the 
time since a bias step of magnitude $\Delta E$. The sample had been 
cooled from above 122~K and held at 25~K for at least 1200~s when the 
bias was stepped. Data points are best fits over the interval 
6~s$<t<360$~s. 
$\blacktriangledown$, $\bullet$, $\blacktriangle$ (for $\epsilon^\prime$ 
data), 
$\triangledown$, $\circ$, $\vartriangle$ (for 
$\epsilon^{\prime\prime}$ data) 
respectively $f$~= 120~Hz, 1.2~kHz, and 12~kHz. All 
$d\epsilon^\prime/d\ln t$ and $d\epsilon^{\prime\prime}/d\ln t$ 
values were negative, as in curve (a) of Fig.~\ref{fig1}. The fitting 
curves are defined in the text (Sec.~\protect\ref{behavior}).}
\label{fig4}
\end{figure}

Figure~\ref{fig4} shows $-s^\prime$ and $-s^{\prime\prime}$ vs. the 
magnitude $\Delta E$ of the bias step. Each pair of 
data points corresponds to a first bias step after a thermal cycle 
to $T_c$ and back to 25~K. $s^\prime$ and $s^{\prime\prime}$ are the 
slopes of the best logarithmic fits to $\epsilon^\prime(t)$ curves as 
in Fig.~\ref{fig3} and corresponding 
$\epsilon^{\prime\prime}(t)$ data for $t$ between 6 and 360~s. 
It is already clear from 
Fig.~\ref{fig3} that $s^\prime$ was not always proportional to 
$\Delta E$. Figure~\ref{fig4} shows that $s^\prime\propto 
s^{\prime\prime}\propto\Delta E$ only at small values. At larger 
$\Delta E$ values $s^\prime$ and $s^{\prime\prime}$ reached limits. 
The limiting value of $-s^\prime$ will be written $s^\prime_0$, 
and another independent scaling 
parameter $\Delta E^\prime_0$ will be defined by putting 
$s^\prime/s^\prime_0=-\Delta E/
\Delta E^\prime_0$ for small $\Delta E$. $s^{\prime\prime}_0$ and 
$\Delta E^{\prime\prime}_0$ similarly define the scale of the 
$s^{\prime\prime}(\Delta E)$ curve. The values of the four parameters 
for each of the three $f$ values of Fig.~\ref{fig4} are given in 
Table~\ref{tab1}. Their precise absolute values are not significant on 
account of the incomplete and nonuniform penetration of the applied 
bias fields into the grains of KDP. It is more useful to note that 
$s^\prime_0$ and $s^{\prime\prime}_0$ are decreasing functions of 
$f$, while $\Delta E^\prime_0$ and $\Delta E^{\prime\prime}_0$ are 
increasing functions. Also, at given $f$, $\Delta E^\prime_0\approx 
1.5\Delta E^{\prime\prime}_0$.

\vbox{
\begin{table}
\caption{Scaling parameters for the rate of change of permittivity of 
a typical pressed powder (Sample~1) at 25~K 
measured between 6 and 360~s after a step, $\Delta E$ of bias 
field. $f$ is the frequency of the low-amplitude measuring field. 
The rates of change are expressed as $s^\prime=t{\rm d}\epsilon^\prime/
{\rm d}t$, $s^{\prime\prime}=t{\rm d}\epsilon^{\prime\prime}/{\rm 
d}t$. Small steps, $\Delta E$ caused proportionate changes, 
$s^\prime/s^\prime_0=-\Delta E/\Delta E^\prime_0$, 
$s^{\prime\prime}/s^{\prime\prime}_0=-\Delta E/\Delta 
E_0^{\prime\prime}$, but $-s^\prime_0$ and $-s^{\prime\prime}_0$ 
represent limiting values for $s^\prime$ and $s^{\prime\prime}$. See 
also Fig.~\ref{fig4}.}
\begin{tabular}{ccccc}
$f$ (kHz) & $\Delta E^\prime_0$ (kV/m) & 
$s^\prime_0$ & $\Delta E^{\prime\prime}_0$ (kV/m) & 
$s^{\prime\prime}_0$\\
0.12 &   31 &  0.0077  &  20 & 0.0018  \\
1.2  &   37 &  0.0055  &  24 & 0.0013  \\
12   &   43 &  0.0039  &  28 & 0.0009\\
\end{tabular}
\label{tab1}
\end{table}}

The curves in Fig.~\ref{fig4} drawn to fit $s^\prime$ at 120~Hz and 
12~kHz correspond to an empirical formula 
$s^\prime/s^\prime_0=-[1+(\Delta E^\prime_0/\Delta E)^2]^{-1/2}$ with 
the $s^\prime_0$ and $\Delta E^\prime_0$ values given in 
Table~\ref{tab1}. The four other curves do not correspond to 
Table~\ref{tab1}, but were obtained from these two by supposing that 
for any given $\Delta E$ value, $s^\prime$ depends on the frequency of 
the measuring field according to a power law, and the response obeys 
the Kronig-Kramers relation. The appeal to Kronig-Kramers is based 
on the reasoning 
that whereas the response to $\Delta E$ on the time-scale of minutes 
is essentially nonlinear, $\epsilon$ represents an approximately 
linear response to a small field on the millisecond time-scale, and a 
set of $\epsilon(f)$ data provides a ``snapshot'' for given $\Delta E$ 
and $t$. The same is true at $t+\delta t$, and so of the derivative 
$s^\prime-is^{\prime\prime}$. It shows there is a link between the 
observations that $\Delta E^\prime_0$ and $\Delta E^{\prime\prime}_0$ 
increase with $f$, and that $\Delta E^{\prime\prime}_0$ is smaller 
than the corresponding $\Delta E^\prime_0$.

\subsection{$T<25$~K}\label{T25K}

The results reported here were obtained with another typical pressed 
powder (Sample~2) prepared from material that had been recrystallized 
in bidistilled water, the solution microfiltered. Referring to 
Fig.~\ref{fig2}(b), $a/b$~= 130\%, $c\approx 0.028$.

\begin{figure}
\centerline{\epsfxsize=7.5 cm \epsfbox{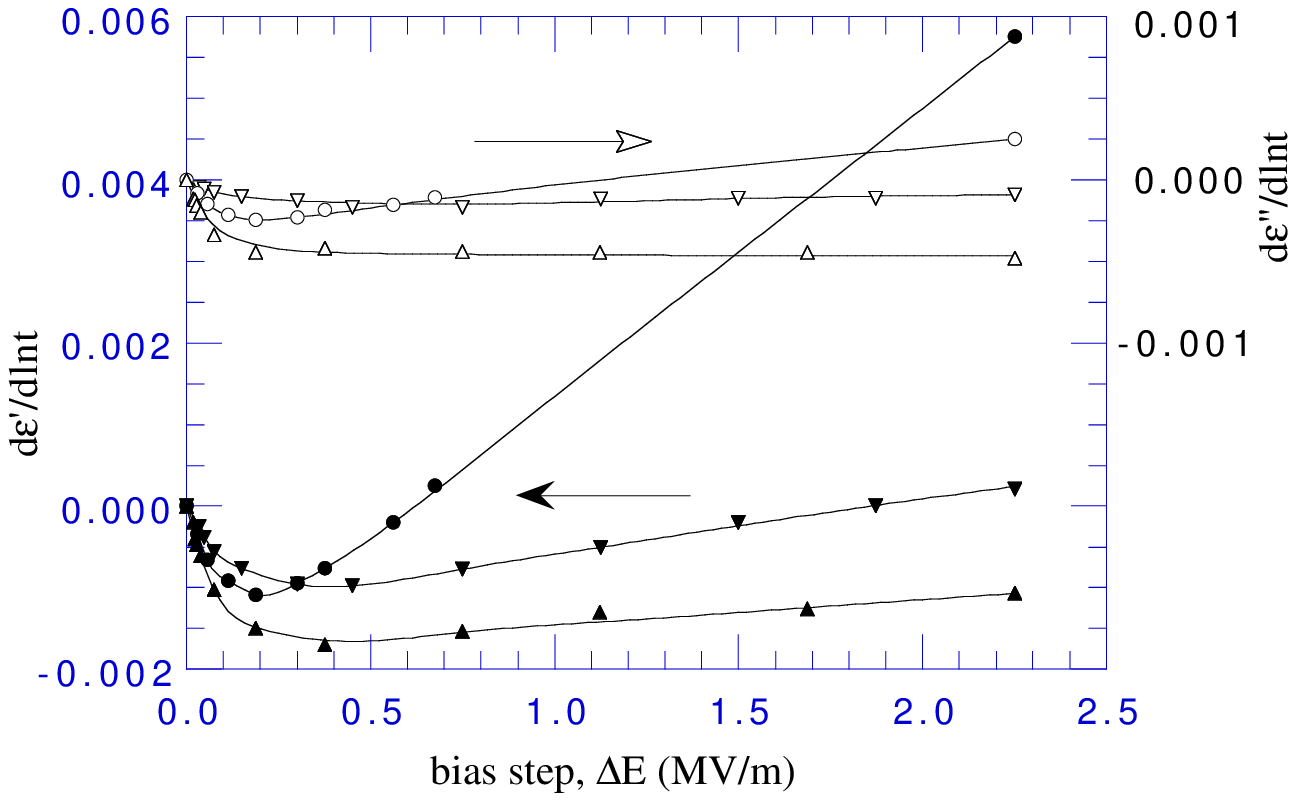}}
\caption{Data for a typical pressed powder (Sample~2) in weak 1.2~kHz 
field: derivatives of real and imaginary permittivity with respect to 
$\ln t$ where $t$ is the time since a bias step of magnitude $\Delta 
E$. Data ponts are best fits over the interval 10~s$<t<$100~s. 
$\blacktriangledown$, $\bullet$, 
$\blacktriangle$ (for $\epsilon^\prime$), $\triangledown$, $\circ$, 
$\vartriangle$ (for $\epsilon^{\prime\prime}$) respectively, $T$~= 4.9, 
7.5, and 12.4~K. Note that unlike Fig.~\ref{fig4} both scales are 
linear and that the scale of $d\epsilon^{\prime\prime}/d\ln t$ 
is expanded and shifted.}
\label{fig5}
\end{figure}

\begin{figure}
\centerline{\epsfxsize=7.5 cm \epsfbox{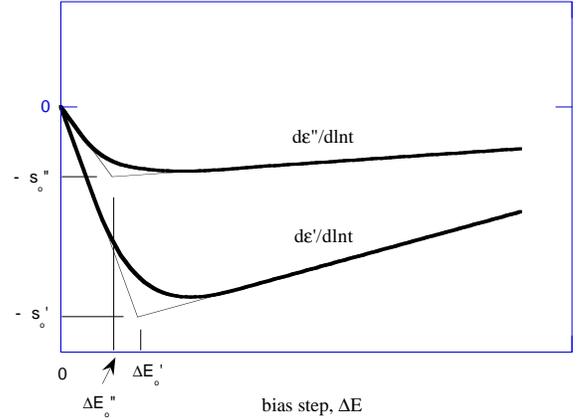}}
\caption{Schematic representation of data as in Fig.~\ref{fig5} to 
show the general definitions of the scaling parameters. In case of 
a ``normal after-effect'' with no ``anomalous 
after-effect'', the rectilinear parts at high $\Delta E$ would be 
horizontal.}
\label{fig6}
\end{figure}

Figure~\ref{fig5} shows $s^\prime(\Delta E)$ and 
$s^{\prime\prime}(\Delta E)$ for Sample~2 at a fixed $f$, fixed $t$ 
interval and three 
temperatures. In this linear plot the details near the origin are not 
clearly seen, but the curves have two straight-line 
sections joined by a curved section. In a similar linear plot the 
same would be true of the 25~K data of Fig.~\ref{fig4}, but at these lower $T$ values 
the straight-line section representing $s^\prime(\Delta E)$ at high 
$\Delta E$ is not horizontal. It has a distinct positive slope at 
each $T$, steep at 7.5~K. In the case of 
$s^{\prime\prime}(\Delta E)$ the effect is much less pronounced. 
It is necessary to generalize the 
definitions of the scaling parameters introduced in 
Sec.~\ref{behavior} and it will be defined that the 
straight lines intersect at $(\Delta E^\prime_0,-s^\prime_0)$ and 
$(\Delta E^{\prime\prime}_0,-s^{\prime\prime}_0)$ respectively as 
shown in Fig.~\ref{fig6}.

\begin{figure}
\centerline{\epsfxsize=8.5 cm \epsfbox{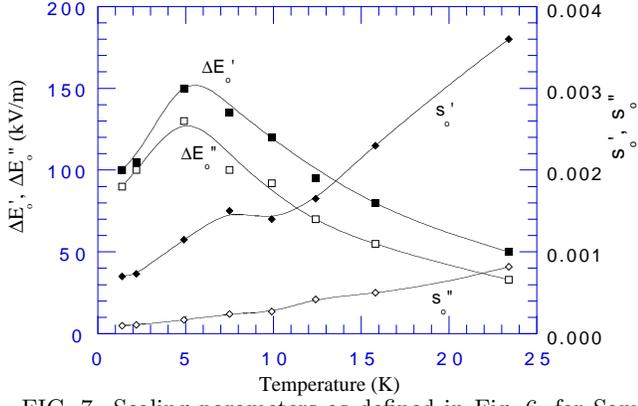}}
\caption{Scaling parameters as defined in Fig.~\ref{fig6}, for 
Sample~2, using a weak 1.2~kHz measuring field. The parameters are 
primarily related to the normal after-effect, but the values obtained 
for $s^\prime_0$ near 7.5~K may have been influenced by the anomalous 
effect. Note that for disordered dielectrics at $T<1$~K, $\Delta 
E^\prime_0\propto T$, while $s^\prime_0$ sharply decreased with 
$T$.\protect\cite{SalvinoPRL94,Saint-PaulJPC86}}
\label{fig7}
\end{figure}

The parameters so defined for Sample~2 at various $T$ are shown in 
Fig.~\ref{fig7}. $s^\prime_0$ and $s^{\prime\prime}_0$ are 
increasing functions of $T$, but $\Delta E^\prime_0$ and $\Delta 
E^{\prime\prime}_0$ peak near 5~K. Below 5~K, $s^\prime_0$ and $\Delta 
E^\prime_0$, $s^{\prime\prime}_0$ and $\Delta E^{\prime\prime}_0$ vary 
in the same proportions, because the response to a small $\Delta E$ 
became temperature independent.

In a linear plot like Fig.~\ref{fig5}, the slopes of the straight 
lines at high $\Delta E$ can be expressed 
in dimensionless units by using $s^\prime_0$ and $\Delta E^\prime_0$ 
or $s^{\prime\prime}_0$ and $\Delta E_0^{\prime\prime}$. 
In such units the initial slope is always -1 by definition. The values 
obtained for these positive slopes of $s^\prime(\Delta E)$ were 0.01, 
0.02, 0.086,0.32, 0.04, and 0.017 respectively at $T$~= 1.37, 2.17, 
4.9, 7.5, 9.9, and 12.4~K. For $s^{\prime\prime}(\Delta E)$ the slopes 
were 0.04, 0.09, 0.003, and 0.000 respectively at $T$~= 4.9, 7.5, 
9.9, and 12.4~K. A closer scrutiny of the 25~K data showed a 
significant positive slope there also, for $s^\prime$ at $\Delta 
E>\Delta E^\prime_0$. Statistical treatment of all data extending 
to $\Delta E>20\Delta E^\prime_0$ (five pressed powder samples) yielded a 
dimensionless slope 0.003$\pm$0.001.

\begin{figure}
\centerline{\epsfxsize=7.5 cm \epsfbox{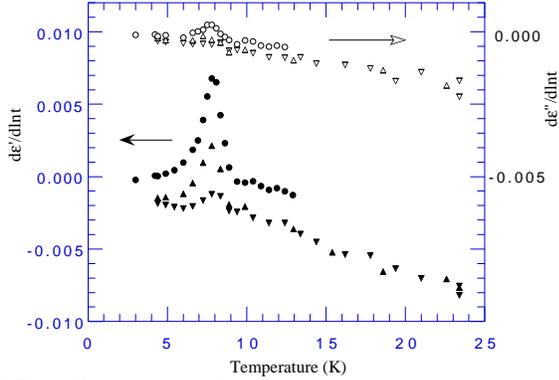}}
\caption{Derivatives of real and imaginary permittivity (in weak 
1.2~kHz fields) with respect to $\ln t$ where $t$ is the time since a 
bias step of fixed magnitude, $\Delta E$, applied at different $T$. 
Data points are best fits over the interval 10~s$<t<100$~s. 
$\bullet$, $\blacktriangle$, $\blacktriangledown$ (for 
$\epsilon^\prime$), $\circ$, $\triangle$, $\triangledown$ (for 
$\epsilon^{\prime\prime}$) respectively Sample~2 with $\Delta E$~= 
2.25~MV/m, another similar sample with $\Delta E$~= 1.32, 0.66~MV/m. Note 
that the scale of $d\epsilon^{\prime\prime}/d\ln t$ is expanded and 
shifted. $d\epsilon^\prime/d\ln t$ has a negative term, that varies 
roughly as $T$ (``the normal after-effect'') and a positive term that 
peaks at $T\approx 7.5$~K (``the anomalous after-effect'').} 
\label{fig8}
\end{figure}

Figure~\ref{fig8} shows $s^\prime(T)$ and $s^{\prime\prime}(T)$, at 
fixed $f$ and fixed $t$ interval, for 
Sample~2 and another similar sample at three different $\Delta E$ values. 
The $\Delta E$ 
values are such that $\Delta E>\Delta E^\prime_0(T)$ always, so that 
far away from 7.5~K, $s^\prime\approx -s^\prime_0$ and 
$s^{\prime\prime}\approx -s^{\prime\prime}_0$. The position of the 
positive peak is independent of $\Delta E$. In relative as well as in 
absolute value it is weaker in $s^{\prime\prime}$ than in $s^\prime$. 

Figure~\ref{fig9} shows $s^\prime(T)$ in the peak region 
at two frequencies and two time lapse intervals. The position depends 
distinctly on the time lapse but not at all on $f$. The magnitude is 
a very slowly diminishing function of $f$, which is consistent with 
the weakness of the $s^{\prime\prime}$ peak (Kronig Kramers).

Figure~\ref{fig10} illustrates another method of studying the upward 
relaxation effect. Two data sets are shown, using different 
procedures. In one case all the $\epsilon^\prime$ measurements were 
made at 4.9~K, where the change $\Delta\epsilon^\prime$ caused by the 
bias step could be measured with little ambiguity because $s^\prime$ 
was relatively very small. The sample was then annealed at 5.45~K for 
2~mn and returned to 4.9~K for another $\epsilon^\prime$ measurement 
without further bias change, annealed at 6.0~K, at 6.6~K and so on. 
The $x$ coordinate in Fig.~\ref{fig10} is the highest anneal temperature 
prior to each measurement. The steepest slope near 7~K corresponds to 
the $s^\prime$ peaks in Figs.~\ref{fig8} and \ref{fig9}. Anneals up to 
20~K continued 
to have some effect, but then a plateau extended to 40~K, at which 
$\sim 75$\% of the $\epsilon^\prime$ shift had been annealed out. 
The original $\epsilon^\prime$ value was almost entirely restored by 
an anneal to 65~K. For the other data set, $\epsilon^\prime$ was 
always measured at 9.9~K, where the bias step was effected, but after 
each 2~mn anneal at different $T$, and after returning to 9.9~K the 
sample was cycled to 122~K. Where they can be compared, ie $T>9.9$~K 
the curves are similar apart from a scaling factor. The scaling factor 
suggests bias step $\Delta E$ was more effective at 9.9~K than at 
4.9~K, doubtless because $\Delta E/\Delta E^\prime_0$ was larger.

A similar procedure was used to compare the strength of the anomalous 
after-effect of different samples after bias steps of different 
magnitudes. The bias was stepped at 6.0~K and the 
sample was annealed for 3~mn at 9.0~K before remeasuring $\epsilon$ at 
6.0~K. This anneal generally restored $\epsilon^\prime$ half way back 
to its original value. The $\Delta\epsilon^\prime$ recorded in these 
cases (and plotted in Fig.~\ref{fig11}) was the change caused by the 
anneal alone, without reference to the original value before the bias 
step.

\subsection{Particle size and impurity effects}\label{particlesize}

The variations of the scaling parameters from one sample to another 
are reported here, and also of the KMN dispersion strength and the 
strength of the upward relaxation of $\epsilon^\prime$ corresponding 
to the anomalous after-effect. The principal results are also 
summarized in Table~\ref{tab2}.

\widetext
\begin{table}
\caption{Summary of the effects of smaller particle size and of added 
impurities compared with ``standard, pure'' pressed powder samples. 
Different impurity species all had qualitatively similar effects.}
\begin{tabular}{p{6cm}p{6cm}p{3cm}}
& Smaller particle size & Added impurities\\
$s^\prime_0$, $s^{\prime\prime}_0$ & little changed 
(slightly~increased) & unchanged\\
$\Delta E^\prime_0$, $\Delta E^{\prime\prime}_0$ & increased & 
increased\\
KMN dispersion & unchanged & weakened\\
strength of anomalous after-effect & unchanged (at~equivalent \hfill{\eject} $\Delta 
E/\Delta E^\prime_0$~values) & weakened\\
\end{tabular}
\label{tab2}
\end{table}
\narrowtext

Firstly it is necessary to consider variations amongst nominally pure 
samples prepared as in Sec.~\ref{measurements}. This category includes 
Samples~1 and 2 that were said to be ``typical''. It was found that 
$s^\prime_0$, $\Delta E^\prime_0$, $s^{\prime\prime}_0$, and $\Delta 
E^{\prime\prime}_0$ for such samples might vary by as much as a 
factor 2, but often much less. This would be due to accidental 
variations of density, homogeneity, and particle size distribution of 
the pressed  powders. Therefore when comparing different categories 
of samples, any variations of these parameters exceeding a factor 2 
are considered significant. Neither 
the scaling parameters nor the KMN strength depended significantly on 
whether the material was used as received or recrystallized, and 
whether in bidistilled water (solution microfiltered) or in deionized 
water. There was no apparent difference between moderately pure and 
highly pure samples. Mean values at $T$~= 25~K and $f$~= 
1.2~kHz were $s^\prime_0\approx 0.005$ and $\Delta E^\prime_0\approx 
35$~kV/m. Referring to Fig.~\ref{fig2}(b), the KMN dispersion strength 
was characterized by 125\%$<a/b<$145\% and 0.025$<c<$0.045.

To investigate the effects of particle size, two samples were prepared 
from powders more thoroughly ground than usual, one of ``as received'' 
material, the other recrystallized. Examination showed many particles 
of globular shape and diameter $\approx 1~\mu$m. With the smaller 
average particle size, these samples undoubtedly contained a higher 
proportion of severely damaged and nonferroelectric material. Both 
had slightly higher $s^\prime_0$ and $s^{\prime\prime}_0$ values, but 
markedly higher $\Delta E^\prime_0$ and $\Delta E^{\prime\prime}_0$ 
values than standard samples. At $T$~= 25~K and $f$~= 1.2~kHz, 
$s^\prime_0\approx 0.007$ and $\Delta E^\prime_0\approx 170$~kV/m. On 
the other hand two samples with larger than normal particles (loose 
powder contained angular shaped particles of dimensions $\approx 
10~\mu$m) yielded similar parameter values as standard samples. The 
larger particles would be likely to have broken up during pressing. 
Similarly, a normally ground sample pressed at 500~MPa had $\Delta 
E^\prime_0$ and $\Delta E^{\prime\prime}_0$ values typical of the more 
thoroughly ground samples pressed as usual at 50~MPa. The KMN 
dispersion strength was found not to depend significantly on particle 
size.

A series of samples was prepared in the usual way from material 
recrystallized from nonstoichiometric or impure solutions. Whatever 
the impurity species, the results were abnormally high $\Delta 
E^\prime_0$ and $\Delta E^{\prime\prime}_0$ values, unchanged 
$s^\prime_0$ or $s^{\prime\prime}_0$ and weakened KMN dispersions. In 
particular, when the solution contained 0.2~H$_3$PO$_4$, 0.1~KHSO$_4$, 
0.2~NH$_4$H$_2$PO$_4$ or 2~RbH$_2$PO$_4$ per 100~KDP, or 2~D$_2$O per 
98~H$_2$O (2\%~d), $s^\prime_0$ ranged from 0.0035 to 0.0077 as usual, 
but $\Delta E^\prime_0$ from 85 to 315~kV/m (always at $T$~= 25~K 
and $f$~= 1.2~kHz) while, referring to Fig.~\ref{fig2}(b), 
101\%$<a/b<$108\% and 0.014$<c<$0.018. Increased impurity 
concentration (0.5~H$_3$PO$_4$, 0.3~KHSO$_4$ or 11~RbH$_2$PO$_4$ per 
100~KDP) caused further increase of $\Delta E^\prime_0$, still no change 
to $s^\prime_0$ but further weakening or disappearance of the KMN 
dispersion. Between 5~K and 25~K the temperature variations of the 
four parameters were roughly similar for pure or impure samples with 
coarse or fine grains (as Fig.~\ref{fig7}).

The impure samples also had weaker anomalous after-effects. 
Figure~\ref{fig11} demonstrates a correlation between strength of 
anomalous after-effect and KMN dispersion strength. Two groups of 
samples are featured. The first group comprises seven nominally pure 
samples. These were prepared as usual (Sec.~\ref{measurements}) or 
were more thoroughly ground than usual so they had a variety of 
$\Delta E^\prime_0$ values, but all had normal KMN dispersions 
($a/b>$125\% and $c>0.025$). The second group included the five impure 
samples mentioned above, and one other that fell into the same 
category (101\%$<a/b<$108\% and 0.014$<c<$0.018). As mentioned in 
Sec.~\ref{T25K}, the strength of the anomalous after-effect was 
characterized by $\Delta\epsilon^\prime$, the change in 
$\epsilon^\prime$ caused by anneal at 9.0~K, following a bias step at 
6.0~K. To allow for variation of sample densities and 
homogeneities, $\Delta\epsilon^\prime$ was normalized with respect to 
$s^\prime_0$ (as measured at $T\approx 25$~K and $f$~= 1.2~kHz). If 
only the ``normal KMN'' samples are selected, 
$\Delta\epsilon^\prime/s^\prime_0$ shows a strong linear correlation 
with $\Delta E$, but only if the latter is also normalized with 
respect to $\Delta E^\prime_0$. The line does not pass through the 
origin, but below it. For $\Delta E<6\Delta E^\prime_0$ ($\Delta 
E^\prime_0$ at 25~K, 1.2~kHz), or equivalently $\Delta E<2\Delta 
E^\prime_0$ ($\Delta E^\prime_0$ at 7.5~K), the normal after-effect 
(downward relaxation) still dominated. The points in Fig.~\ref{fig11} 
for the ``weak KMN'' samples fall near another line of lower slope, 
indicating a weaker anomalous after-effect.

One other impure sample deserves a special comment. Following a known 
example,\cite{NakamuraJJAP81,AbeJPSJ84} KDP was recrystallized from 
solution containing a large excess of base, 58~KOH for 100~KDP. The 
crystals were very hygroscopic and when removed from the dessicator 
and pressed into pills, fluid was expelled and filled the space 
between grains (pill density = 96\% of crystal density). Several 
other samples were more or less conductive at room temperature, but 
in this respect, this one was an extreme case. Nevertheless, it 
behaved at low $T$ almost as a usual, slightly impure sample. At 
25~K, $\epsilon^\prime$, $\epsilon^{\prime\prime}$, $s^\prime_0$ and 
$s^{\prime\prime}_0$ were rather higher than usual. The $ac$ field 
would have penetrated the grains more effectively, on account of the 
higher permittivity of the intergranular space. The KMN dispersion 
was observed near 12~K (at 1.2~kHz) as usual, as also the anomalous 
after-effect centered around 7.5~K. This demonstrates conclusively 
than none of the reported low $T$ effects is a purely surface effect, 
and that the KMN dispersion is unlikely to be a Maxwell-Wagner 
effect, as already argued by Kuramoto {\it et al.}\cite{KuramotoJPSJ84}

The 2\% deuterated sample was classed amongst the impure samples 
because of its properties outlined above. It is logical to suppose 
that protonic impurity in KD$_2$PO$_4$ (DKDP) would have a similar 
effect, so that if an analog to the KMN dispersion occurs in DKDP it 
would be necessary look for it in a $d>98$\% sample. Such an effect 
was searched at 5~K$<T<$80~K but not found with a commercial 
(Aldrich) 98\% $d$ sample nor with another KDP sample recrystallized 
twice from 99.8\% D$_2$O in an atmosphere free of natural humidity.

\section{Single crystals}\label{singlecrystals}

Four single-crystal samples were studied. SC~1 consisted of two 
slabs, cut normal to the $c$-axis and silver electroded by evaporation: 
total area 72~mm$^2$, mean thickness 0.73~mm. The two pieces were 
connected in parallel. SC~2 consisted of two other $c$-cut slabs, 
gold electroded, also connected in parallel: 38~mm$^2\times 0.35$~mm. 
SC~3 idem but the two pieces were cut normal to an 
$a$-axis and silver painted: 75~mm$^2\times 0.94$~mm. SC~4 was a single 
$c$-cut plate of irregular shape 61~mm$^2\times 0.30$~mm, gold 
electroded.

Both the normal (at 25~K) and the anomalous (near 7.5~K) after-effects 
were observed with SC~1 and SC~2 ($E\parallel c$). All low $T$ 
dielectric properties were sensitive to cooling speed through $T_c$, 
and whether cooled in field or no field. They were less accurately 
reproducible from one thermal cycle to another than with pressed 
powders. With SC~1, after fast zero-field 
cooling, $\vert s^\prime\vert$ and $\vert s^{\prime\prime}\vert$ were 
typically several times smaller than with pressed powders,
\cite{GilchristSSC98} but so 
also were $\epsilon^{\prime\prime}$ and $d\epsilon^\prime/dT$. At 
25~K, attempts to apply fields $>200$~kV/m always caused an 
instability. At lesser $\Delta E$ values, $\vert s^\prime\vert$ varied 
roughly as $\Delta E^{0.4}$, and no $s^\prime_0$ or $\Delta 
E^\prime_0$ value could be estimated. Near 7.5~K, the anomalous 
after-effect was characterized by a lower $\alpha$ constant in the 
stretched exponential law. Thus if 
$\epsilon^\prime(t)=\epsilon^\prime(\infty)-C\exp-(t/\tau)^\alpha$, 
$\alpha$ took the value 0.45 for pressed powder Sample~1 at 7.5~K, 
but 0.34 for SC~1 at the same $T$. With SC~2, also fast 
zero-field cooled, all the low $T$ dielectric properties were several 
times weaker than with SC~1.

With SC~3 very little after-effect was found following a bias 
step, $\Delta E$~= 300~kV/m at 25~K. This puts an upper limit to 
$\vert s^\prime\vert$ of $1.5\times 10^{-4}$, so if any effect exists 
for $E\parallel a$ it is of a lower order of magnitude than for 
$E\parallel c$. The inference is that the after-effects of the 
pressed powders are essentially caused by the $E\parallel c$ field 
component.

SC~4 was used to check the Arrhenius law of the KMN-C absorption. 
For this purpose the calibration of the carbon resistor thermometer 
was not relied on, but a germanium resistor was placed in close 
thermal contact with the sample. As KMN noted,\cite{KuramotoJPSJ84} 
the background absorption and dispersion that has to be subtracted 
depends on $f$ as well as $T$. Some plausible choice has to be made, 
how this is done. KMN assumed at each $T$, a Cole-Cole law plus a 
linear $\epsilon^{\prime\prime}(\log f)$ background. For the present, 
the background to be subtracted at each $T$ and $f$ was assumed to be a 
linear interpolation of the data at $T$~= 6.14 and 19.75~K (i.e. well 
below and above the KMN effect) and the same $f$. This also 
subtracted an instrumental 
and circuit error that becomes serious at $f>10$~kHz, and it resulted 
in a symmetric $\epsilon^{\prime\prime}(\log f)$. The 
$\epsilon^{\prime\prime}$ peak position was found at each $T$. Two 
field values were used, 3.5 and 7.0~kV/m (cf.~1.0~kV/m~
\cite{KuramotoJPSJ84}) and the 
data sets analysed separately. The 3.5~kV/m results are shown in 
Fig.~\ref{fig12} and yielded Arrhenius parameters $f_0$~= 173~GHz, 
$A$~= 20.09~meV. The 7.0~kV/m results would not be distinguishable in 
the plot and gave $f_0$~= 170~GHz, $A$~= 20.03~meV. If the 6.14~K 
data alone had been subtracted as background, the figures would have 
been $f_0$~= 
605(640)~GHz, $A$~= 21.65(21.68)~meV. Assuming the linear 
interpolation is more 
appropriate, the present result is not significantly different from 
the published\cite{KuramotoJPSJ84} result and it may be concluded that 
$A=19.5\pm 0.7$~meV and 40~GHz~$<f_0<200$~GHz.

Also shown in Fig.~\ref{fig12} are analogous results for pressed powder 
Sample~2, at 3.8~kV/m, using the same background subtraction. It might 
be expected that $\epsilon^{\prime\prime}(f)$ at 6~K would have a peak 
near 100~Hz corresponding to the $E\parallel a$ dispersion,\cite{KuramotoJPSJ84} and 
so give a false baseline. Such an effect was searched between 5 and 
7~K but not detected, which means it could have been no more than 3\% 
as strong as the absorption and dispersion near 12~K in this sample. 
Best Arrhenius parameters for the latter were $f_0$~= 880~GHz, and 
$A$~= 21.16~meV, and it is concluded from its position in 
Fig.~\ref{fig12}, and absence of $E\parallel a$ effect (at 5--7~K) 
that it corresponds almost entirely to the KMN-C effect.

Figure~\ref{fig12} also shows two points representing the time constant 
of the anomalous after-effect. The time constant $\tau(T)$ was taken 
to equal $t$ at the point of maximum positive slope of 
$\epsilon^\prime(\log t)$. It was found at $T$~= 
7.44 and 7.82~K, and ($2\pi\tau)^{-1}$ is plotted. The points fall 
distinctly below the Arrhenius law, whichever of the three data sets 
is extrapolated, but the law derived from the same pressed powder 
sample comes nearest. If an estimated normal after-effect contribution 
of the form $Ct^{-p}$ had been subtracted from the data, the 
discrepancy would have been smaller.

\section{Discussion}\label{discussion}

The after-effect in KDP at $T<25$~K consists of two parts that are 
readily distinguished. Both parts have been reported for single 
crystal samples,\cite{ZimmerFL87,GilchristSSC98} as well as pressed 
powders and the KMN-C dispersion\cite{MotegiJPSJ83,KuramotoJPSJ84} is 
also a single-crystal property. The discussion of their origins must 
be broad enough to encompass both types of sample.

\subsection{The normal after-effect}\label{normalafter-effect}

A first salient feature of the normal after-effect is that it has been 
observed at $T$ ranging from 1.4 to 300~K, so that its $T$ range 
extends well above $T_c$ and well below 
the usual domain freezing temperature in crystals ($\sim~100$~K). A 
second is that it has no characteristic relaxation time, and a third 
is that its magnitude reaches a limit ($s^\prime_0$, 
$s^{\prime\prime}_0$) at very moderate values ($\Delta E^\prime_0$, 
$\Delta E^{\prime\prime}_0$) of applied field step, with pressed 
powders at least. 
Below 25~K, it conforms to a general law for ``glassy'' 
properties in that $\epsilon$ is a simple, 
near-linear function of $T\log t$.\cite{PrejeanJP80} The low $\Delta 
E^\prime_0$ and $\Delta E^{\prime\prime}_0$ values that decrease with $T$ 
suggest a high dipole moment value that increases with $T$. All this suggests 
microdomains, that were present accidentally in the single crystals, but 
more systematically present in the powders. The notion of microdomains was 
developed\cite{YokosukaJJAP86} to explain the diffuse nature of the phase 
transition in disordered ferroelectrics such as PLZT ceramics. 
The sizes and shapes are widely distributed. R. H\"ohler {\it et al.}
\cite{HohlerPRB91} invoked this notion to explain their 
finding of a retarded dielectric response at low $T$ in a PLZT ceramic. 
Although a different $\epsilon$ was measured, related to the polarisation 
and not the $ac$ polarisability, a $T\log t$ law was also found, 
between 20 and 80~K, in conditions corresponding to $s^\prime=-s^\prime_0$ 
(retarded response independent of 
$\Delta E$). This possibly suggested independent thermally activated Debye 
processes with a uniform distribution of activation energies, which would 
imply a uniform spectral density of $\log\tau$ values ($\tau$ is a 
relaxation time) that varies as $T$. Strongly interacting systems that 
relaxed according to a hierarchical sequence\cite{PalmerPRL84} would also 
have been possible, and more plausible. The authors 
suggested\cite{HohlerPRB91,HohlerJNCS91} a discrimination in favour 
of the latter, based on results of applying a $T$ step as well as a 
$\Delta E$ pulse, which indicated an 
abnormally low frequency prefactor. For the present, it is also more 
plausible to suppose strongly interacting systems in the pressed powder 
samples. This would explain why $s^\prime_0$ and $s^{\prime\prime}_0$ 
always took similar values at a given $T$. In the single crystals 
there may not have been enough active microdomains to reach the 
strong-interaction limit. Apparently $\Delta E^\prime_0$ and $\Delta 
E^{\prime\prime}_0$ represent a threshold bias-step value for a 
prompt and widespread 
microdomain polarization rearrangement that limits the extent to which 
the system can be out of equilibrium. It is also most likely that 
microdomains are responsible for the ``background'' dispersion and 
absorption at low $T$, both in single crystals ($E\parallel c$) and 
powders. They also explain naturally why $\Delta E^\prime_0$ and 
$\Delta E^{\prime\prime}_0$ decreased with $T$, but increased with $f$. 
Either increased $T$ or decreased $f$ would bring into play larger 
microdomains.

The microdomains that are active at low $T$ are probably ones that 
have less than full orthorhombic distortion, so they are easily 
switched. In single crystals they are likely to be associated with 
crystal defects, in pressed powders likely to be much influenced by 
the damaged grain surface layers. In the powders they must be subject 
to a very broad distribution of stresses, as witnessed by the 
equivalence of field-cooled and zero-field-cooled properties. 
On the other hand, the active microdomains at ambient $T$ 
are perhaps to be identified with the orthorhombic inclusions that 
have been reported in the tetragonal phase.\cite{SuvorovaSPC91}

A ``normal'' after-effect of a bias step is also characteristic of 
disordered 
dielectrics generally at very low $T$. The immediate rise of 
$\epsilon^\prime$ and $\epsilon^{\prime\prime}$ followed by decay as 
$\log t$ was observed with hydroxyl-doped KCl at $T<1$~K,
\cite{Saint-PaulJPC86} and several structural glasses, also at 
$T<1$~K.\cite{SalvinoPRL94} The results could be displayed in plots 
like the present Figs.~\ref{fig3} and \ref{fig4}, though the samples were not 
thermally cycled between measurements as in the present study. The 
behaviour of the structural glasses below 1~K was 
explained\cite{CarruzzoPRB94} by reference to a random Ising model of 
dipoles with long-range interactions.\cite{KirkpatrickSSC78} One difference 
from the present results 
was that the $\Delta E^\prime_0$ parameter was an increasing function of $T$. 
In accordance with the model, it was found $\Delta E^\prime_0\approx kT/p$, 
where $p$ is the (fixed) relevant dipole moment. Another difference 
was that $s^\prime_0$ decreased sharply with $T$. The other main point in 
common is that $\epsilon^\prime$ and $\epsilon^{\prime\prime}$ relaxed 
downwards, and this can be understood as the system 
of interacting dipoles gradually self-trapping into deeper potential wells, 
from which it can respond less actively to weak applied fields.

\subsection{The anomalous after-effect}\label{anormalafter-effect}

The most obvious distinguishing features are that $\epsilon^\prime$ 
relaxed upwards, not 
down and that although some effect was detectable at all $T$ between 1.4 and 
25~K, there was a pronounced peak near 7 or 8~K, depending on the time lapse. 
Also, the anomalous effect was only observed when $\Delta E>\Delta E^\prime_0$, 
and then its strength as represented for example in Fig.~\ref{fig11} 
increased as $(\Delta E/\Delta E^\prime_0-a)$, where 
$a\approx 1$. The $\Delta E^\prime_0$ parameter needed to describe 
the normal after-effect appears as a threshold for the anomalous 
effect. More precisely there was a tendency for the strength 
to level off and possibly saturate at some high $\Delta E$ value not 
reached in this work.\cite{GilchristSSC98} This suggests that the unit 
dipole moment involved 
here is much smaller than the moments of the microdomains. Since the 
anomalous after effect is also linked to KMN-C, it may be 
supposed a same species of dipole (``the KMN dipoles'') is responsible.

A phenomenological model can then be formulated 
as follows. Two subsystems (the microdomains and the KMN dipoles) 
coexist and interact, each having its own relaxation dynamics. The KMN 
dipoles only interact weakly with one another, but certainly 
interact with the microdomain system. After a bias step that is big enough 
to unsettle the microdomain system, this settles into a new 
configuration that is metastable, with the KMN dipoles in their present 
states. These dipoles then relax with their characteristic time constant, 
$\tau$, perpetually changing the local strain fields and electric 
fields. This affects the microdomain system qualitatively as would a 
random series of applied field changes. Referring to the effect of 
repeated bias changes that was mentioned in 
Sec.~\ref{switches}, and summing over the sample volume, it might be 
expected that after a single applied field step, $\Delta E>\Delta 
E^\prime_0$,
\begin{eqnarray*}
\epsilon^\prime(t)=\epsilon^\prime(\infty)+Ce^{-t/\tau}t^{-p}+
\frac{C^\prime}{\tau}\int^t_0e^{-(t-t_i)/\tau}(t-t_i)^{-p}dt_i
\end{eqnarray*}
For simplicity, unstretched exponentials have been written here. 
The term in $C$ represents regions of the sample where 
$\epsilon^\prime(t)$ is still relaxing downwards due to the applied 
field step. Its volume diminishes exponentially. The $C^\prime$ term 
represents the sum of micro-regions that have been affected by a 
subsequent KMN dipole flip, that 
occurred at $t_i$. The expression allows an upward relaxation of 
$\epsilon^\prime(t)$ only if $C^\prime>C$. This is possible because 
if $\Delta E>\Delta E^\prime_0$, $C$ takes its limiting value related 
to $s^\prime_0$, while $C^\prime$ can plausibly exceed this limit. The 
field changes caused by the dipole flips may be very strong, but they 
are localized and random, so they do not cause a prompt and widespread 
microdomain rearrangement like a strong applied field change. The 
microdomain system can therefore be driven further away from 
equilibrium.

\subsection{Possible origin of KMN-C and anomalous 
after-effects}\label{KMN-C}

KMN suggested\cite{KuramotoJPSJ84} that a peculiar mode of motion 
related to the domain wall structure may be responsible. If some of 
the atoms located at the domain boundary were in shallow double 
potential wells, movement of these atoms across the barriers would be 
a possible origin. This brings to mind the hydrogen atoms and their 
O\ldots H-O bonds. At the time of KMN, the potential barrier to 
H-bond reversal was thought to be well over 
100~meV,\cite{LawrenceJPC80} but newer 
work\cite{SugimotoPRL91,Yamada94,YamadaJPSJ94,IkedaPB96} based on 
neutron incoherent scattering data has yielded the much lower estimate 
of 37.1~meV,\cite{Yamada94} and again a precise potential function 
with a barrier height near 30~meV (for DKDP, 135~meV) has been 
calculated for the system comprising a hydrogen atom and coupled 
lattice mode, based on 30~K data.\cite{IkedaPB96} The principal aim 
was to account for $T_c$ and its deuteration shift by considering 
lattice dynamics at $T>T_c$, but the result might also apply to a 
special case where a single H bond reversal could occur at low $T$ 
without the energy cost of creating a HPO$^{2-}_4$ ion and an 
adjacent H$_3$PO$_4$ group.

It is necessary to examine possible domain-wall structures. Wall width 
increases with $T$ towards $T_c$,\cite{AndrewsJPC86} but at $T\ll T_c$, 
walls are plausibly approximated by the vanishingly thin models 
proposed by Barkla and 
Finlayson.\cite{BarklaPM53} There were two such structures, the 
``polarized'' and the ``neutral'' domain wall. Both respect the 
Slater rule (no HPO$_4$ and no H$_3$PO$_4$ groups). It is 
not known for certain which is the more accurate approximation 
to real walls at 
$T\ll T_c$ but Bjorkstam and Oettel calculated that the polarized wall 
would be the more stable.\cite{Bjorkstam66} Bornarel\cite{BornarelJAP72} 
showed that either of these 
walls, normally planar and perpendicular to an $a$-axis, can have any 
number of lateral step displacements. The minimum displacement is a 
half lattice parameter. The step displacements, or quasi-dislocations, 
like the planar walls, respect the Slater rule, but only if they run 
straight across the entire crystal, parallel to the $c$-axis. They are 
associated with an intense local strain field. Minimum domain wall 
movement therefore involves glide of a quasi-dislocation along its 
entire length, or the presence of HPO$_4$ and H$_3$PO$_4$ groups. 
Otherwise, if the walls have finite width, a minimum movement without 
HPO$_4$ or H$_3$PO$_4$ groups consists of six simultaneous H-bond 
reversals.\cite{Schmidtref34} The simplest case with HPO$_4$ or 
H$_3$PO$_4$ is a unit jog, where a quasidislocation shifts by one 
lattice parameter. This requires a single HPO$_4$ or H$_3$PO$_4$ group 
and its illustration\cite{BornarelF75} is reproduced in 
Fig.~\ref{fig13} for the case of a polarized 
domain wall. The case of a neutral wall is almost entirely equivalent. 
By a sequence of single H-bond reversals, the vacancy (or the excess 
H) can move from one PO$_4$ to the next within a $c$-stack of unit 
cells, and so, in principle right across the crystal, together with 
the associated jog. In practice it may get pinned somewhere along the 
line, by a crystal dislocation or an impurity. A few such pinned jogs 
may persist as the material is cooled down to temperatures where 
their creation by thermal activation would be virtually impossible. 
Also, progression of a jog requires successive reversals of 
differently oriented H-bonds, first one that lies near the plane of 
the domain wall, then one nearly perpendicular to it and so on. It is 
plausible that sometimes, because of the local fields and strains one 
of these, but not the other remains possible at low $T$, because in 
just one case the two states are near energetically equivalent. This 
would constitute the double potential well system.

Yamada and Ikeda\cite{Yamada94} considered the ``cluster tunneling 
mode'' or protonic polaron\cite{YamadaJCSJ98} model, but found that 
their purpose of predicting hydrogen dynamics at $T>T_c$ was better 
served by a model of incoherent tunneling between self-trapped states. 
At low $T$ this would become coherent phonon-assisted tunneling, with 
an expected transition probability $\propto T^7$. On the other hand, 
the protonic polaron would have an extremely small tunneling 
splitting, so that thermal activation down to 12~K, or even 7~K, 
would be more plausible. Moreover, the tunneling mode would be 
overdamped at higher $T$, and in that condition the predictions of the 
two models would be experimentally indistinguishable.

In another development since KMN, evidence has been 
reported\cite{MeloBJP92} of an orthorhombic-monoclinic phase 
transition in KDP near 60~K. This would mean the H-bonds within a 
ferroelectric domain are not all equivalent, as previously supposed, 
but not enough is known about the new phase to draw any other 
conclusions.

Meanwhile, a different origin for the KMN-C effect cannot be totally 
rejected. This would attribute it to a defect species that is 
intrinsic to solution grown KDP (for example a growth dislocation or 
an included water molecule). At low $T$ this defect would only be 
dielectrically activated by the presence of a domain wall.

\section{Conclusions}\label{conclusions}

A new dielectric property of ferroelectric KDP has been reported, 
``the anomalous after-effect''. It has been shown to be related to 
the ``normal'' after-effect\cite{ZimmerFL87} and also to the KMN-C 
dispersion.\cite{MotegiJPSJ83,KuramotoJPSJ84} The normal after-effect 
is a well-known property of $c$-cut KDP crystals, and of many other 
ferroelectric materials. Like the $T^{3/2}$ specific heat term in 
microcrystalline KDP, it is reminiscent of such an effect in 
structural and dipole glasses, but differs in certain important details. 
It is attributable to microdomains with a wide distribution of sizes, 
shapes and stresses. Study of the normal after-effect with pressed 
powders has allowed a set of scaling parameters to be determined for 
a series of samples of different qualities. As shown in 
Fig.~\ref{fig11}, these scaling parameters also apply to the anomalous 
after-effect, which therefore also involves microdomains.

The KMN-C dispersion is another known property of $c$-cut KDP crystals. 
Such dispersions are generally caused by point defects in crystals, 
and are unknown in structural glasses at low temperatures. Before the 
present work there was no apparent link between KMN-C and the normal 
after-effect. The KMN-C dispersion possibly owes its origin to rare, 
isolated HPO$_4$ and H$_3$PO$_4$ groups associated with jogs on 
lateral steps (quasidislocations) of domain walls. The activation 
energy of 19~meV, which is also shared by the anomalous after-effect, 
would then be related to the energy barrier for the reversal of a 
single H-bond in this particular environment. It is not clear how 
closely this should be assimilated to the hypothetical barrier for 
the reversal of a single H-bond in tetragonal KDP, that has been the 
object of recent estimates.\cite{Yamada94,IkedaPB96} A correlation 
might be expected, if different ferroelectric compounds of the KDP 
type could be compared. The most interesting comparison would be with 
DKDP, but so far, as mentioned in Sec.~\ref{particlesize} no 
dispersion analogous to the KMN effect has been observed with DKDP. 
For the other isostructural compounds the situation can be summarized 
as follows. As pressed powders, RbH$_2$PO$_4$ and the arsenates all 
behaved analogously to KDP, but the arsenates had first to be 
crystallized with excess base (solution pH$>$6). Each compound exhibited 
a low-field dispersion that obeyed an Arrhenius law, and a 
corresponding anomalous after-effect. With RbH$_2$PO$_4$ the 
activation energy was close to the value for KDP, perhaps 1~meV 
higher. With KH$_2$AsO$_4$ it was near 30~meV, with CsH$_2$AsO$_4$, 
44~meV and with RbH$_2$AsO$_4$ between these two. However, no 
independent estimates of the barrier heights are available.

The anomalous after-effect results from an interaction between two 
coexisting subsystems. One subsystem relaxes with a definite 
relaxation time, the other with a very broad distribution of 
relaxation times. The result of the interaction is an apparent 
tendency of the system to evolve temporarily away from its stable 
equilibrium. The lines of an explanation sketched in 
Sec.~\ref{anormalafter-effect} need to be developed into a model.

\acknowledgments
This work owes much to Jean~Bornarel, who supplied copious advice and 
background knowledge together with the electroded single crystals. 
Comments by Jean~Souletie, and information supplied by Andr\'e~Durif 
and Marie-Th\'er\`ese~Averbuch were also appreciated.

\newpage

\newpage

\newpage

\vbox{
\begin{figure}[t]
\centerline{\epsfxsize=7.5 cm \epsfbox{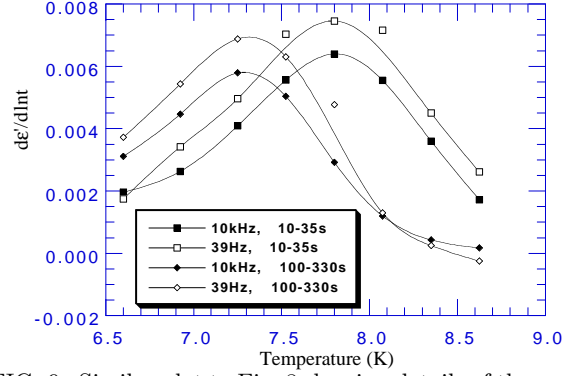}}
\caption{Similar plot to Fig.~\ref{fig8} showing details of the 
anomalous positive peak. All data here are for Sample~2 in 3.8~kV/m 
fields of two different frequencies, $f$ following bias steps of 
magnitude $\Delta E$~= 2.25~MV/m. $f$~= 39~Hz ($\square$, 
$\lozenge$), 10~kHz ($\blacksquare$, $\blacklozenge$). Data points are 
best fits over the interval 10~s$<t<35$~s ($\blacksquare$, $\square$), 
or 100~s$<t<330$~s ($\blacklozenge$, $\lozenge$). 
$d\epsilon^\prime/d\ln t$ reaches a peak at a temperature that 
depends on $t$, not on $f$.} 
\label{fig9}
\end{figure}}

\begin{figure}
\centerline{\epsfxsize=7.5 cm \epsfbox{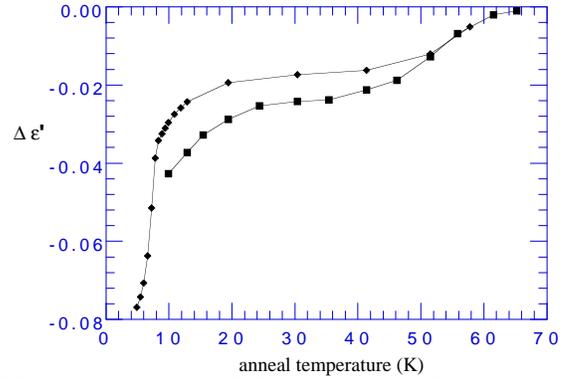}}
\caption{Anneal-out of the change in real permittivity of Sample~2 
(in weak 1.2~kHz field) caused by a bias step of 2.25~MV/m at 
4.9~K ($\blacklozenge$) or 9.9~K ($\blacksquare$). Different 
procedures were used as explained in the text (Sec.~\ref{T25K}) but 
$\epsilon^\prime$ 
was always measured at 4.9~K or 9.9~K, never at the anneal 
temperature and 
$\Delta\epsilon^\prime$ is the change in $\epsilon^\prime$ resulting 
from the bias step and any subsequent annealing. The anneal 
temperature is either the temperature of a single 2~mn anneal 
($\blacksquare$) or the highest, and latest of a series of 2~mn 
anneals ($\blacklozenge$).}
\label{fig10}
\end{figure}

\begin{figure}
\centerline{\epsfxsize=7.5 cm \epsfbox{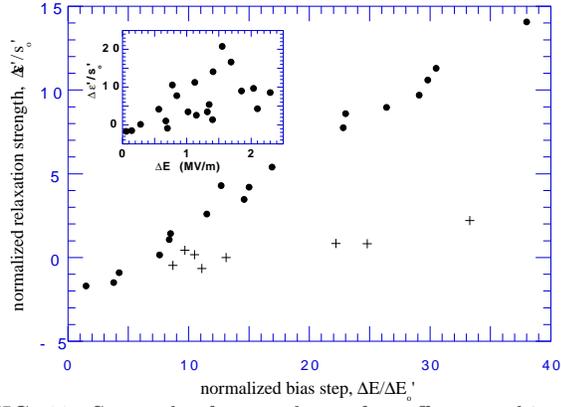}}
\caption{Strength of anomalous after-effect vs. bias step 
magnitude for thirteen pressed powder samples. The bias was stepped at 6.0~K, 
the sample was annealed for 
180~s at 9.0~K then returned to 6.0~K. $\Delta\epsilon^\prime$ is the 
change in real permittivity (weak 1.2~kHz field, at 6.0~K) caused by 
the anneal. Where the normal after-effect was dominant it is 
negative, where the anomalous effect dominated, positive. 
$\Delta\epsilon^\prime$ is plotted normalized with respect to 
the scaling parameter $s_0^\prime$ (at 25~K, 1.2~kHz). The 
samples form two groups: pure samples with KMN dispersions of normal 
strength ($\bullet$), 
impure samples with weak KMN dispersions ($+$). In the main plot, 
the bias step magnitude is also normalized with respect to $\Delta 
E^\prime_0$ (25~K, 1.2~kHz). If it is not (inset, points $\bullet$ only) no 
clear pattern emerges. This is a key result. The anomalous 
after-effect is seen to correlate with the KMN dispersion, but also to 
be linked to the normal after-effect since the same scaling 
parameter, $\Delta E^\prime_0$ is involved.}
\label{fig11}
\end{figure}

\vbox{
\begin{figure}
\centerline{\epsfxsize=7.5 cm \epsfbox{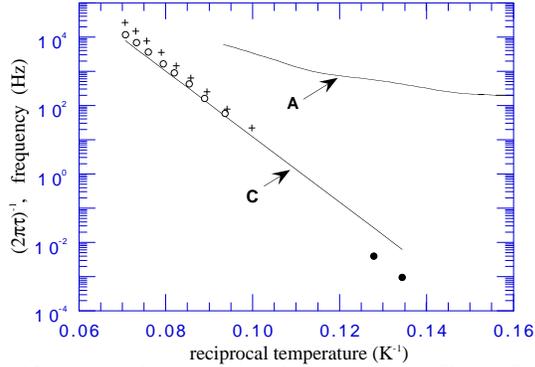}}
\caption{Arrhenius plot showing data from Single Crystal Sample 
SC~4, Pressed Powder Sample~2 and from Kuramoto {\it et al.}
\cite{KuramotoJPSJ84} $\circ$, frequency 
of absorption peak of SC~4 at 3.5~kV/m, with background subtracted 
as stated in text; $+$, idem of Pressed Powder Sample 2 at 
3.8~kV/m; $\bullet$, ($2\pi\tau)^{-1}$ of Pressed Powder Sample~2, where $\tau$ is 
the value of $t$ at the inflexion point (maximum positive slope) on 
the plot of $\epsilon^\prime$ vs. $\log t$ after a bias step of 
2.25~MV/m. The line labeled C is the Arrhenius fit given by 
KMN,\cite{KuramotoJPSJ84} and the curve labeled A an 
approximate fit to their $E\parallel a$ data.}
\label{fig12}
\end{figure}}

\begin{figure}
\centerline{\epsfxsize=5.5 cm \epsfbox{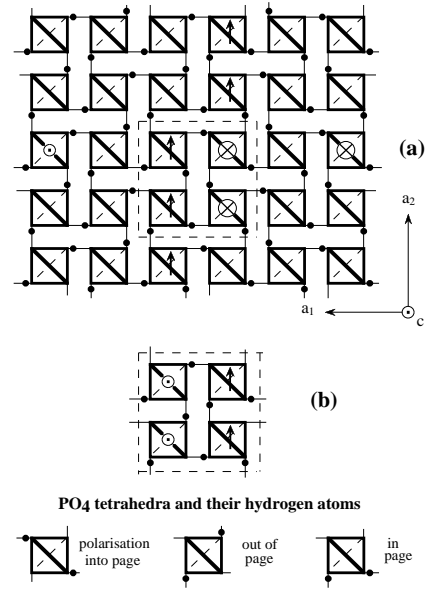}}
\caption{(a)~Conventional
\cite{BarklaPM53,Bjorkstam66,BornarelJAP72,Schmidtref34,BornarelF75} 
representation of KH$_2$PO$_4$ showing a polarized domain wall with a 
lateral step displacement. 
Black dots represent H atoms. The dashed line encloses a stack of 
(tetragonal) unit cells that lies astride the domain boundary. Within 
this stack, the PO$_4$ tetrahedra lie on a helix, whose pitch is the 
lattice parameter. (b) the same stack viewed at a higher level. To 
pass from one arrangement to the other, any one of the tetrahedra 
within the stack must possess only one closely bound H (the ion is 
HPO$_4^{2-}$). The fault can travel up or 
down the helix by a series of single H-bond reversals as explained by 
Bornarel.\cite{BornarelF75} An opposite jog 
would require one H$_3$PO$_4$, and would be similarly mobile.}
\label{fig13}
\end{figure}

\widetext

\end{document}